# WORKPLACE DIVERSITY AND INNOVATION PERFORMANCE: CURRENT STATE OF AFFAIRS AND FUTURE DIRECTIONS

**Christian R. Østergaard**
Aalborg University Business School
**Bram Timmermans**
NHH Norwegian School of Economics

***Abstract***

*When we talk about diversity in the field of innovation studies, we tend to take a strong position on the underlying knowledge this diversity represents. Here, the main argument is that it is the diversity in knowledge available to and subsequently recombined by innovative agents that are a predictor of both the ability to generate innovation and to how radical such innovations are. The diversity of knowledge has been studied from several angles, ranging from the diversities in a firm's technology, product, or patent portfolio to diversity in firms' collaborations with innovation partners. In the last 10 years, we have observed an ever-increasing interest in a specific form: diversity in human capital. Fueled by the business case for diversity, there is an interest in understanding how the combination of people with different backgrounds fosters the innovation performance of firms. Studies have measured diversity on a wide range of personal-level characteristics, at different levels of the organization, and in particular kind of settings. Innovation performance has been measured using an arsenal of indicators, often drawing on a large range of databases. This chapter takes stock of this research, identifying the current state of affairs and proposing future research trajectories in the field of diversity and innovation.*

## Introduction

Innovation is generally considered to be the recombination of existing knowledge and ideas (Schumpeter, 1942; Nelson and Winter, 1982; Henderson and Clark, 1990). Not surprisingly, the ability of a firm to bring together different forms or a diversity of knowledge and ideas is regarded as an important determinant of its innovative capabilities. The field of innovation studies has studied the recombination argument by understanding the depth and breadth of a firm's knowledge base through analyzing the portfolio of products (Pavitt, 1998), patents (Breschi et al., 2003), and technologies (Suzuki and Kodama, 2004). With the growing interest in open innovation, we also observe an increasing focus on the diversity of interaction with external parties like customers, suppliers, universities, and other external organizations (Laursen and Salter, 2006).

While there is clearly a longstanding interest in understanding the impact of diversity on innovative capabilities, the focus on how diversity in employee composition contributes to innovation performance is a more recent phenomenon. While a few studies on this topic did exist (e.g., Bantel and Jackson, 1989), it was not until the early 2010s that a larger body of empirical research began to emerge, in which different aspects of diversity in the workforce, teams, and boards and their relation to a wide variety of innovation performance indicators were tested. This research stream builds on a longstanding and rich tradition of studying the role of diversity in organizations and its relation to organizational performance in sociology, management, and organization studies; see reviews on the topic by Williams and O'Reilly (1998) and Van Knippenberg and Schippers (2007). This literature is marked by a general interest in understanding how organizational performance is affected by the diversity of workers in terms of achieved and ascribed personal characteristics that include gender, age, race, cultural background, nationality, education, and work experiences. Individuals differ along many of these dimensions simultaneously, but the literature has focused on measurable characteristics, especially demographic characteristics, and often examined one diversity dimension at a time. Typically, these different dimensions are treated as unit-level compositional constructs that sort people into groups based on a common attribute (Harrison and Klein, 2007). This means that diversity is understood as the distribution of differences according to various attributes or dimensions.

While workplace diversity is regarded as an important driver for innovation, it is generally difficult to empirically investigate this claim. Information about firms' innovation activities and outcomes is required, as is access to detailed data on the composition of employees at the appropriate levels of the firm. These two data types need to be combined to empirically analyze the relation between workplace diversity and innovation. Access to both types of data has proven to be challenging, but during the last decade a growing number of studies have analyzed the relation between diversity in boards, teams, and workplaces and innovation performance in many industries and countries, using a wide variety of data and diversity indicators.

The purpose of this chapter is thus twofold. On the one hand, we present an overview of existing empirical research on the relation between workplace diversity and innovation performance. We present the overall findings of the relevant studies, emphasizing the theoretical framework, unit of analysis, dimensions of diversity, and underlying mechanism(s) that each study uses, along with empirical issues. At the same time, we use our examination of existing empirical research to identify gaps in the literature and thus avenues for future research that can reveal whether, under which conditions, and to what extent workplace diversity can actually contribute to increased innovative capabilities.

This chapter is structured as follows. In the next section, we focus on the different theoretical perspectives that guide our intuitions about the relation between workplace diversity and innovation performance. We then begin our review by identifying the relevant empirical literature, elaborating on how we define workplace diversity and innovation performance and presenting the review's inclusion and exclusion criteria. In the next section, we classify the articles and present the current state of research with synopses of each paper. The chapter concludes by suggesting directions for future research.

## Theoretical Perspectives on Workplace Diversity and Innovation

Workplace diversity typically refers to individual-level differences that exist within an organizational setting, which can be the organization as a whole or a particular unit, like a top management team, board, or R&D department. While such differences might be ascribed to any number of individual attributes, existing research tends to focus on those that are relatively easy to identify or obtain, such as demographic differences like age, gender, and ethnic and cultural background or qualifications like tenure, education, and functional background (Williams and O'Reilly, 1998; van Knippenberg and Schippers, 2007).

In investigating the impact of diversity, we seek to understand how the perception that other individuals in the organization are different and the underlying variety in knowledge, skills, and capabilities associated with different attributes affects worker's attitudes, organizational behavior, and organizational performance. Depending on the attitudes, processes, and performance metrics of interest, workplace diversity might make very different contributions. This also explains why it is challenging to arrive at any conclusive findings on the role of workplace diversity and why it is often characterized as a double-edged sword (Milliken and Martins, 1996; Williams and O'Reilly, 1998).

To understand how workplace diversity affects the performance metric of innovation, we take how innovations emerge as our point of departure. There is certainly not a single path, as sources of innovation are plentiful (von Hippel 1988; Drucker, 1998; Tidd and Bessant, 2020). Innovations tend to emerge when an organization combines previously disconnected knowledge and ideas (Schumpeter, 1942; Nelson and Winter, 1982; Henderson and Clark, 1990), which can occur accidentally or as the result of a deliberate process. It involves the recombination of existing knowledge from different sources within an organization and identifying and capturing knowledge from outside the organization (Tidd and Bessant, 2020). The process might be driven by a specific innovation agenda, a reaction to a recognized market need, or a response to sheer necessity (Tidd and Bessant, 2020). To understand how workplace diversity affects innovation performance of firms, we must first grasp how diversity contributes to any innovation drivers. Fortunately, research on workplace diversity and innovation has been able to build on a well-established academic literature that guides our intuition. This literature can be roughly divided into two camps.

The first is research that emphasizes the value of the variety in knowledge resources that underlie workplace diversity. More specifically, this literature stream regards individuals as unique resources that differ in education, knowledge, information, experiences, competencies, skills, and cognition. These knowledge resources can contribute in different ways to innovation performance, either directly through immediate recombination of these knowledge sources and the ability to recognize and act on opportunities or through mobilizing resources that are not currently under the organization's control. Thus, workplace diversity facilitates learning, information flows, knowledge sharing, problem solving, and knowledge recombination (Cohen and Levinthal, 1990; Nooteboom, 2000; Wenger, 2000; van der Vegt and Janssen, 2003; Østergaard et al., 2011). There is a non-exhaustive list of research strands that could be included under this heading, but they share a strong position in strategic management and the behavioral theory of the firm, which includes evolutionary theories and organizational learning, all of which provide arguments for the importance of access to a variety of input factors to introduce (successful) innovations.

One of the more dominant perspectives applied in research on diversity and innovation is resource theories, including the resource-based view (Penrose, 1959; Barney 1991) and the knowledge-based view (Kogut and Zander, 2003; Grant, 1996; Richard, 2000; Felin and Hesterly, 2007), which take an explicit stand on how the unique combination of resources and capabilities that can emerge from a diverse workplace provides firms with long-term competitive advantages. Research on dynamic capabilities (Teece et al., 1997; Eisenhardt and Martin, 2000; Teece, 2007; Roberson et al., 2017) links diversity to the ability to sense, seize, and transform an organization. Upper-echelon theories (e.g., Hambrick and Mason, 1984; Simons et al., 1999) make related claims, offering strong arguments for how cognitive diversity in top management teams allows those leadership groups to consider alternative strategic options.

Another set of resource theories feeds our intuition as to how diversity affects innovation and behavioral theories of the firm (Cyert and March, 1963), primarily the subfields of organizational learning (Nooteboom, 2000; Wenger, 2000) and evolutionary economics (Nelson and Winter 1982; Dosi, 1988). Both views highlight the importance of search processes to secure a variety of knowledge that leads to new combinations of what is known. Such search processes are often highly local and close to the firm's existing knowledge base; however, while local searches are often efficient, they might not be enough to address a particular need (Laursen, 2012). Furthermore, local searches might lack access to the wide variety of knowledge necessary for new recombinations (Laursen, 2012). Workplace diversity opens up the search space for an organization both via the broader range of knowledge that individuals in the organization possess and, as research on open innovation states, because workplace diversity allows an organization to access a variety of external knowledge sources (Cohen and Levinthal, 1990; Bogers et al., 2018).

While the resource camp emphasizes the positive effects of diversity on innovation performance, it also acknowledges certain limitations. Nooteboom (2000) has observed that there is a higher scope of interactive learning when two pieces of knowledge are cognitively close. These claims are very much aligned with organizational limits on absorptive capacity; that is, the ability of a firm to recognize knowledge, assimilate it, and apply it to commercial ends (Cohen and Levinthal, 1990).

Rather than focusing on the value of diverse resources in themselves, the second camp emphasizes how the attitudes of workers, categorization, and subsequent group processes are affected by workplace diversity. These arguments emerge from the field of social psychology and refer more specifically to processes of social identify theory (Tajfel, 1981), social categorization (Turner, 1987), similarity attraction (Byrne, 1971), and unconscious bias (Greenwald and Banaji, 1995). These perspectives are closely linked to perceptions of diversity, the subsequent attitudes of workers, and how workers then behave toward one another. When confronted with (perceived) differences between (groups of) individuals in the workplace, many people have a fundamental desire to distinguish themselves from others into separate groups, often based on differences in individual attributes. While categorization is not inherently problematic, it can create adverse interaction effects. Due to in-group and out-group membership, a positive bias might be shown to in-group members, with out-group members regarded as less attractive, trustworthy, honest, or cooperative, which can eventually lead to conflict (Joshi and Jackson, 2003).

Similarity attraction is clearly related to self-categorization but is not driven by it. Here, it refers to the interpersonal attraction that arises due to similarities between the various members of a group; the idiom is "birds of a feather flock together" (McPherson, et al., 2001). This attraction is a result of shared experiences and values that ease communication and interaction between members and enhance their cohesiveness. However, this might lead to a dislike of members of other groups (Horwitz, 2005).

Both views highlight that the presence of workplace diversity can create barriers that prevent the recombination of knowledge, not that workplace diversity cannot contribute to organizational performance in general or innovation in particular. Research in this camp has demonstrated that diversity does indeed add new perspectives, bring in different ideas, and increase awareness and assessment of problems and knowledge that are less familiar to the dominant group. In that regard, it reduces problems of groupthink (Janis, 1971). Thus, it is critical to understand how workplace diversity affects the use of information (Dahlin et al., 2005), where different diversity constructs might face challenges of "being heard," to acknowledge that the size distribution of the different groups that emerge plays a role (Milliken and Martins, 1996), and to recognize the fact that barriers resulting from social categorization and similarity attraction might decrease over time (Horwitz, 2005).

To summarize, we can identify two camps that offer predictions on the relation between workplace diversity and innovation, but we want to emphasize that they are not polar opposites; rather, they operate in tandem. As innovation is the outcome recombining existing knowledge, the presence of a diverse workforce – and therefore a diverse knowledge base – would increase the likelihood of an organization's introducing an innovation. However, the mere presence of diversity might not be sufficient to promote innovation if attitudes and work processes prevent this diversity in knowledge from being recombined. Diversity in the workplace does not always equal inclusion. If social identity theory, social categorization, and similarity attraction are indeed fundamental processes of how people interact on the work floor or in the office, management practices and work organization are moderators that either reinforce or counteract those phenomena.

## Identifying the Relevant Empirical Literature

To investigate the current state of affairs on the relation between diversity and innovation performance, we conducted a systematic literature review. Before we elaborate on our sampling procedure, we first discuss our inclusion criteria. This review focuses on empirical evidence on the relation between workplace diversity and innovation performance, which means we exclude conceptual papers and reviews on the topic. Furthermore, the primary subject of an article had to address diversity and innovation performance. A lack of conceptual discussion and cases where diversity is included only as a control variable were excluded from the sample (e.g., the share of STEM workers or foreign workers in an organization).

Our interest in workplace diversity also means that we required diversity to be measured among internal stakeholders in the organization, whether that means the organization as a whole or one or more of its subunits. In addition, we only included studies where workplace diversity is an objective and quantifiable measure of diversity, such as age, experience, gender, and nationality, along with a broad range of other types of diversity (see Tables 1 and 2). Consequently, studies where management provides a subjective perception of the level of diversity in the workplace (e.g., Bouncken et al., 2016) fall outside our selection criteria. This was done to avoid relying on subjective perceptions of diversity that could be subject to managers' unconscious bias in assessing both visible and non-visible types of diversity. Management might think that a team or workplace is truly diverse, while an objective measurement would reveal that most members share the same characteristics. This criterion also eliminates studies that only investigate diversity through dummy variables, such as the presence of females or foreign nationals in a team.

In this chapter, we apply a narrow definition of innovation performance. More specifically, we refer to the output of an innovative process as a manifestation of a new idea, the registration of a patent, the announcement of a new product, or the introduction of new products, services, and processes (Freeman and Soete, 1997; Hagedoorn and Cloodt, 2003). This is a broader definition of innovation than used in the Oslo Manual (OECD/EUROSTAT, 2018), since it includes other types of innovation indicators like patent-based indicators and composite measures that include innovation, such as a combination of R&D budget, new product development, and technological innovation (Wei and Lau, 2012), and idiosyncratic survey measures. Tables 1 and 2 provide a non-exclusive overview of the wide range of innovation types included.

The definition of innovation excludes other types of performance- and innovation-related indicators, such as R&D spending, creativity, innovation strategy, openness in collaboration on innovation, innovation climate, networks, knowledge sourcing, and simulation models. In scanning the literature on workplace diversity and innovation, we observe many articles that link workplace diversity to organizational behavior and practices that are associated with innovation performance but do not investigate this relation directly. Examples include the relation between workplace diversity and R&D investments (Atallah et al., 2020) or innovation activities (Bello-Pinto and Bianchi, 2020). These are all variables where a relation to innovation performance has previously been established, but unless an article explicitly addresses innovative output, it is not included in our analysis. Furthermore, with the rise of open innovation, we also observe an interest in investigating how diversity affects the choice and extent of external collaborators on innovation (e.g., Bogers et al., 2018; Fitjar and Solheim, 2017). While we acknowledge that collaborations, including the depth and breadth of such efforts (Laursen and Salter, 2006), are often a driver for innovation performance, we only include such studies if they also empirically address innovation performance (e.g., Mohammadi et al., 2017).

To identify relevant articles, we conducted a Boolean search of peer-reviewed articles in Scopus, which is a comprehensive but curated scholarly database. We conducted a title, abstract, and keyword search on the word "innovation" combined with "diversity," "heterogeneity," or "related variety," as well as "employee*," "manager*," "team*," or "board*." We limited our search to English language journals in three Scopus-defined fields: Business, Management and Accounting; Economics, Econometrics and Finance; and Social Sciences.[1] We further restricted our search by selecting only journals listed in the *Academic Journal Guide 2018* from the Chartered Association of Business Schools and only those within a field where the topic of innovation and workplace diversity appears on the research agenda: Entrepreneurship and Small Business Management, Human Resource Management and Employment Studies, Innovation, Information Management, Operations and Technology Management, Operations Research and Management Science, Ethics CSR management, Marketing, Social Science, Organizational Studies, and Strategy. A total of 508 journals were included in the search, which yielded 371 articles distributed across 115 journals.

With this list as a point of departure, the authors independently conducted close readings of all titles and abstracts to identify articles that deal explicitly with an empirical analysis of (i) workplace diversity and (ii) innovation performance. When this information could not be deduced from the title or abstract, or when an abstract was not available, we included the paper in the next step of the selection process. After this close reading, the authors compared their lists and took an extra round with articles on which there was initial disagreement to reach a consensus. This process resulted in 99 articles. The next phase was reading the full papers. Similar to the abstract reading, the authors independently assessed whether articles fulfilled our selection criteria, compared their assessments, and discussed articles where there was disagreement. After this procedure, we ended up with a total of 34 articles for detailed analysis.

Before we analyzed these articles, we conducted a backward and forward citation search to identify additional articles on the topic of workplace diversity and innovation performance that were overlooked in our initial selection procedure, including articles published in journals that were not part of the initial selection. Based on those forward and backward citations, we were able to identify eight additional articles.

## The Current State of the Literature

The 42 articles were reviewed using a set of coding categories. First, we focused on the overall perspectives on workplace diversity and innovation performance. Second, we elaborated on the perspective of the unit of analysis; that is, the part of an organization and the organizational members for which a given diversity measure was constructed. Third, we described the type of innovation. Fourth, we identified which individual-level characteristics an article used to measure diversity. Fifth, we elaborated on the nature of the relation between diversity and innovation (i.e., direct or indirect). Sixth, we described the context of the study in more detail. Finally, we crafted an overview of the empirical results of the paper and the extent to which it found workplace diversity to contribute to innovation performance. The authors read and coded the articles independently to bolster the integrity of the coding process; afterward, the authors compared the individual assessments and reached agreement on the coding of all articles.

[1] We excluded Medicine (medi), Agricultural and Biological Sciences (Agri), Dentistry (dent), Chemical Engineering (ceng), Biochemistry, Genetics, and Molecular (BIOC), Materials Science (MATE), Nursing (NURS), Neuroscience (NEUR).

The coding used is presented in Table 1, which lists and numbers all 42 articles. It also details for each paper the unit of analysis, innovation type, diversity type, data type, specific context, nature of the workplace diversity–innovation relation, expected direction of this relation, and overall findings. In Table 2, we have aggregated and classified the articles based on the construct of diversity, unit of analysis, type of innovation, and main results. The numbers in the cells correspond with the paper numbers from Table 1.

*--INSERT TABLE 1 AROUND HERE—*

*--INSERT TABLE 2 AROUND HERE--*

## Perspective and Mechanisms on Workplace Diversity and Innovation performance

Before we dig deeper into the overall characteristics of the articles, we elaborate on the theoretical perspectives applied in them. As noted in the first section of this chapter, workplace diversity and innovation performance can be placed in two broadly defined and interrelated camps: "diversity-as-a-resource" theories and "attitudes-and-group-processes" theories. The identified articles tend to focus on the former, with a particular emphasis on the positive role workplace diversity plays in innovation performance. Most articles thus argue for a positive relation between innovation and the respective diversity constructs. A few studies adopt a more agonistic perspective, arguing for both negative and positive relations (e.g., Bantel and Jackson, 1989; Wei and Lau, 2012; Özgen et al., 2017; Wikhamn and Wikhamn, 2020). Only a few articles explicitly state that a negative relation is anticipated (Østergaard et al., 2011; Galia and Zenou, 2012; Brunetta et al., 2020). In addition, some studies argue that diversity promotes innovation but note that too much diversity could be negative, suggesting a curvilinear or inverted-U relation (Chi et al., 2009; Østergaard et al., 2011; Kim and Kim, 2015; Gonzáles-Morena et al., 2018). Whether this aligns with the actual empirical findings is discussed below.

Most of the studies in our sample argue for a direct relation between workplace diversity and innovation performance, thus focusing on the main effect. A few investigate more explicitly whether workplace diversity acts as a moderator (Moser et al., 2019; Rejeb et al., 2020) or whether workplace diversity indirectly affects innovation (Somech and Drach-Zahavy, 2013; Ruiz-Jiménez et al., 2016). Another set of studies investigates whether the relation between workplace diversity and innovation performance is moderated by other firm characteristics; this includes stage in the organizational life cycle (Tzabbar and Margolis, 2017), ownership (Fernández, 2015), size (Yap et al., 2005), task and goal interdependence (Van der Vegt and Janssen, 2003), strategic consensus (Camelo et al., 2010), team dynamics (Wei and Lau, 2012), or other work processes (Chi et al., 2009; Cheung et al., 2016; Kristinsson et al., 2016; Bocquet et al., 2019; Xie et al., 2020). One study embeds workplace diversity in a regional setting (Lee, 2015). Although they often do so implicitly, these studies build on resource theories but are cautious about the presence of mechanisms explained by the above-mentioned attitude and group processes. A few studies argue that some types of diversity might interact negatively with others (Garcia Martinez et al., 2017; Zouaghi et al., 2020).

## Unit of Analysis

While the articles revolve around workplace diversity, the authors often limit themselves to a subset of workers. An especially salient divide in the line of inquiry is the focus on upper-echelon teams – managers, founders, and owners (e.g., Bantel and Jackson, 1989; Nathan and Lee 2013; Lee 2014) – compared to a broader workforce perspective, although the latter are generally restricted to or emphasize a particular type of worker, such as high-skilled workers (e.g., Mohammadi et al., 2017; Laursen et al., 2020), R&D workers[2] (e.g., Díaz-Garcia et al., 2013; Fernández, 2015; Garcia-Martinez et al., 2017; González-Moreno et al., 2018; Wikhamn and Wikhamn, 2020), top and middle management (Schubert and Tavassoli, 2020), and blue- or white-collar workers (Parrotta et al., 2014). The distinction aligns nearly perfectly with the type of data being used (see Types of Data section below). As Table 2 reveals, there is less empirical evidence on non-upper-echelon teams in research on workplace diversity and innovation.

Articles on the different types of teams tend to focus on the innovation performance for which a specific team bears immediate responsibility, regardless of whether it is innovation performance as a responsibility for top management (Li et al., 2016) or specific innovation output for R&D workers (Chi et al., 2009). Articles that adopt a broader workforce perspective tend to focus on the innovation performance of the organization at large. The main argument of these articles is that innovation performance cannot be solely attributed to a particular group within an organization, but that innovation is an interactive process that is carried out by a larger group of employees in various departments, including workers at lower levels of the hierarchy (e.g., Østergaard et al., 2011). By adopting a whole-workplace perspective, one study argues that interaction and knowledge spillovers between individuals within an organizational unit and across organizational units lay the foundation for the recombination of knowledge to occur. Others take a position on how workers with a particular, often foreign, background contribute to the innovation performance of the company as a whole (Laursen et al., 2020).

Another divide can be drawn based on the type of involvement with the overall innovation process: largely strategic or primarily operational. The majority of articles take a strong strategic position by drawing inspiration from upper-echelon theories (Hambrick and Mason, 1984; Bantel and Jackson, 1989), investigating the diversity in boards of directors (n=4), top management teams (n=11), or founding teams (n=4). These articles argue that diversity in the upper echelons leads to different strategic priorities rather than delivering impact through implementation. Consequently, these articles emphasize a dynamic capability perspective (Schubert and Tavassoli, 2019). Those studies with an operational focus investigate diversity in R&D teams or among R&D and (other) highly skilled workers (n=11). Given their operational focus, they have a strong resource- and knowledge-based perspective, with the variety of knowledge skills and competencies leading to innovation performance.

[2] In several instances, the articles refer to R&D "teams" but actually study all the R&D workers in an organization.

## Construct of Diversity

To assess the extent of diversity in a workforce, it is helpful to consider that the construct of diversity has three dimensions: the number of groups distinguishable by a given attribute, the balance between the various groups, and disparity between the groups (Stirling, 2007). It is important to note that diversity should always be viewed in context; it does not exist in a vacuum. Therefore, what increases diversity at one level of an organization might not increase it at another. In empirical studies of diversity, the disparity dimension is often neglected because it is difficult to assess the degree of differences between groups. As a result, most studies tend to address disparity only indirectly when assigning different attributes to specific groups while treating the disparity between groups as fixed (e.g., Laursen et al. 2020). The articles apply various measures of diversity that can be divided into inherited characteristics (e.g., age, gender or a measure of culture, ethnicity, or nationality) and acquired characteristics (e.g., tenure, functional or occupational background, work and industry experience, and education). Most articles tend to emphasize either inherited or acquired characteristics; only six use a combination of the two.

Table 2 reveals that, among papers that investigate inherited characteristics, understanding the role of gender is the most common topic of interest, followed closely by diversity of cultural or ethnic background, with the latter making an explicit reference to (highly skilled) migrants (Özgen et al., 2013, Nathan and Lee, 2013; Lee 2014; Özgen et al., 2017; Laursen et al., 2020). Age diversity has received less attention, but several studies do investigate or control for average age. These are the characteristics around which most of the literature on the business case for diversity revolves, examining how specific minority or marginalized groups can contribute to organizational performance, regardless of whether this diversity is measured among boards, top management teams, R&D teams, or (highly skilled) workers. There are basically two perspectives from which these studies argue that inherited characteristics can contribute to innovation performance. First, differences in (observable) inherited characteristics are argued to represent underlying differences in skills, knowledge, perspective, and practices. For example, gender diversity leads to broader perspectives and a wider knowledge based (e.g., Díaz-Garcia et al., 2013). Theoretically, these studies align with the resource perspective. The other view focuses on how differences in these characteristics can trigger attitude and group processes, emphasizing barriers to and bottlenecks of diversity and the consequent likelihood of interpersonal conflict between groups. As a result, studies that investigate the role of inherited characteristics tend to have a slightly more balanced view on the possible positive and negative consequences of diversity.

As to acquired characteristics, Table 2 reveals that most articles focus on the role a worker fulfills in the workplace, in terms of both function and occupational background. This type of diversity highlights the need to involve individuals with different skills and the knowledge and competencies linked to their work tasks. In upper-echelon teams, functional diversity makes a distinction between roles like general manager, marketing manager, finance, production and manufacturing, R&D, information technology, and human resources (e.g., Bantel and Jackson, 1989; Camelo et al., 2010; Protogerou et al., 2020). In studies on specific R&D and medical teams, the functions are more specific to a specialized team; for example, researchers, technicians, and auxiliary personnel in R&D teams (Fernández, 2015) and physicians, nurses, social workers, occupational therapists, and dieticians in medical teams (Somech and Drach-Zahavy, 2013). Overall, studies that rely on this type of diversity use the team as the unit of analysis and gather information through surveys of teams. A few of these studies use standardized occupation codes listed in register data (Fernández, 2015). Studies relying on functions and occupational background tend to emphasize the need for the different skills and competencies in these functions for an effective innovation process that can lead to successful innovation performance. However, Cheung et al. (2016) argue that the suggested benefits of diversity in these cross-functional teams are contingent on the presence of trust among members.

In addition to function and occupational background, studies draw on different measures of diversity in experience. Studies measure both diversity in experience within an organization (in the form of tenure; e.g., Bantel and Jackson, 1989; Yap et al., 2005; Chi et al., 2009; Camelo et al., 2010; Wei and Lau, 2012; Li and Huang, 2019) and diversity in general industry experience (Herstad et al., 2019; Li, 2019; Solheim et al., 2018, 2020). Concerning diversity in tenure, the least frequently used diversity type, arguments point to how differences in tenure represent differences in social groups that negatively affect communication (Bantel and Jackson, 1989), while others argue that those differences in perspective could be positive for innovation (Camelo et al., 2010; Li and Huang, 2019), although this relation could be curvilinear (Chi et al., 2009). Industry experience is measured by identifying the career history of workers in the organization (e.g., Herstad et al., 2018). Diversity in this category represents both a resource-based view and a dynamic capability perspective on diversity; consequently, industry diversity is often associated with better innovation performance.

The final acquired characteristic that many studies examine is diversity in educational background (Bantel and Jackson, 1989; Chi et al., 2009; Østergaard et al., 2011; Wei and Lau, 2012; Somech and Drach-Zahavy, 2013; Parrotta et al., 2014; Li et al., 2016; Tzabbar and Margolis, 2017; Garcia Martinez et al., 2017; Mohammadi et al., 2017; Solheim and Herstad, 2018; Li and Huang, 2019; Schubert and Tavassoli, 2020). The argument for focusing on educational diversity is that education provides people with basic concepts and models for problem solving (Østergaard et al., 2011), is important for people's absorptive capacity (Cohen and Levinthal, 1990; Backmann et al., 2015), makes people members of a community of practice (Wenger, 2000), and provides people with a knowledge base. Thus, diversity in education among a firm's employees facilitates potential combinations of different bodies of knowledge, and this form of diversity is unequivocally associated with increasing innovation performance (Østergaard et al., 2011; Mohammadi et al., 2017; Schubert and Tavassoli, 2020), although some argue for a curvilinear relationship (Østergaard et al., 2011).

## Types of Data

Empirically investigating the relation between workplace diversity and innovation performance is difficult since it requires access to detailed information concerning both inherited and acquired individual-level characteristics and firm-level performance indicators. To deal with these challenges, the articles draw on various data sources, often in combination, since some sources of data are suitable for measuring diversity and others innovation performance. The type of data used also aligns with the unit of analysis chosen in a given article.

Articles with a team focus rely on data from tailored surveys that allow for more detailed information on all team members and capture relevant indicators for innovation performance. These surveys restrict themselves to a particular sample of firms, such as publicly traded firms (Kim and Kim, 2015; Li et al., 2016; Rejeb et al., 2020; Li, 2019), firms in a narrowly defined industry of firm population (Bantel and Jackson, 1989; Chi et al., 2009; Somech and Drach-Zahavy, 2013; Kristinsson et al., 2016; Tzabbar and Margolis, 2017). The focus on teams also means selecting multiple teams in one organization (Cheung et al., 2016), which offers the strength of filtering out significant amounts of performance variation. The time-consuming nature of hand-curated datasets and surveys means that many of these studies have a relatively small sample size. Nevertheless, they do have some inherent strengths. By adopting a survey design, such studies are able – beyond examining the direct relations between diversity and innovation performance – to capture additional team processes that might act as moderators and mediators on subsequent performance outcomes. Examples include the extent to which there tends to be strategic consensus (Camelo et al., 2010), modes of interacting and communication (Li et al., 2016), and human resources practices (Chi et al., 2009).

A smaller set of studies tries to investigate workplace diversity by making use of annual reports (Galia and Zenou, 2014; Boone et al., 2019; Li and Huang, 2019) or gaining access to business intelligence databases that have information on management and board members (Kim and Kim, 2015; Li, 2019; Brunetta et al., 2020). These papers tend to focus on the upper echelons of an organization: top management teams and boards. Like tailored surveys, they often rely on a subset of industries. However, unlike survey-driven studies, these sources do not provide information on work processes or direct measures of innovation performance. Consequently, these data are often combined with information from dedicated innovation surveys or by relying on patent statistics.

In the last two decades, the increased availability of linked employer–employee databases has proven a rich source for investigating workplace diversity (Østergaard et al., 2011; Parrotta et al., 2014; Mohammedi et al., 2017; Özgen et al., 2013, 2017; Solheim and Herstad, 2018; Herstad et al., 2019; Brixy et al., 2020, Laursen et al., 2020; Schubert and Tavassoli, 2020; Solheim et al., 2020). Register or register-like data provide access to a detailed overview of inherent and acquired characteristics of all employees and managers in all firms in an economy. Consequently, sample sizes are extremely large and capture the heterogeneity of an entire nation's organizational landscape. This detailed level of information also allows for restricting the analysis to a subset of workers based on, for example, occupation codes and educational backgrounds. The unit of analysis in these studies demonstrates that nearly all of them apply one of these restrictions. An additional benefit is that longitudinal information is available on who works for what firm at a given point in time. This feature allows researchers to measure diversity at different moments in time and provides opportunities to touch on causal relations or apply panel models to investigate the role of diversity. While registers offer some obvious strengths, they also have weaknesses. Unlike studies that rely on teams, little to no information can be obtained on work processes. Furthermore, there are limitations to identifying the tasks of workers, including the extent to which they work in teams, and consequently how much they contribute to overall innovation performance. However, individual workplaces are often relatively small, which mitigates some of these concerns.

Studies that focus solely on surveys tend to include questions designed to reveal the innovation performance of the team or organization. Those relying on annual reports, data from business intelligence firms, and register data often employ innovation surveys and patent statistics. Those that use innovation surveys often choose the Community Innovation Survey (CIS; n=13) (e.g., Mohammadi et al., 2017; Herstad et al., 2019; Schubert and Tavassoli, 2020) or CIS-inspired surveys (e.g., Østergaard et al., 2011; Nathan and Lee 2013; Lee 2014) that try to capture innovation performance as defined by the Oslo Manual. In most countries, these surveys are administered by the national statistical agencies, which means that they have relatively high response rates. So, although business surveys in general suffer from poor response rates, these innovation surveys offer richer data, especially among larger and more innovation-prone organizations across industries. Consequently, sample sizes are often several thousand firms. In addition, since many of these surveys are centrally administered and conducted at regular intervals, there are opportunities to conduct longitudinal studies that link workplace diversity with innovation performance. However, this empirical strategy has thus far rarely been used. In addition to information on innovation activities, many of these surveys have detailed information on innovation-related input measures, including the level of R&D investments and innovation activities in firms, which are relevant controls or possible moderators for innovation performance.

Patent statistics are another source of data (n=6); the surveys apply measures using the number of patents applied for or granted (e.g., Parrotta et al., 2014; Li and Huang, 2019; Boone et al., 2019), devising a weighted innovation performance indicator by taking into consideration the number of forward citations (Laursen et al., 2020) or identifying newness by investigating whether the firm applied for a patent in a new patent class (Li, 2019). While patent statistics are available for all patent-active firms, this automatically results in limiting the analysis to the subset of firms that are active in patenting in industries where patenting is frequent.

## Findings

The key question is simple: What evidence does the empirical literature offer on the relation between workplace diversity and innovation performance? For this purpose, we determine whether the studies find positive, negative, or non-significant relations between workplace diversity and innovation. Table 2 provides a summary that records the findings of the 42 articles, with distinctions made for type of innovation output, diversity construct, and organizational unit under observation.

As noted above, most articles emphasize a positive and direct relation between workplace diversity and innovation performance. Given that workplace diversity is often referred to as a "double-edged sword," some may view this position as overly optimistic. However, the overall results do seem to be warranted, since most studies find a positive relation between diversity and innovation, with only a few finding statistically significant negative relations. Positive relations are reported for several diversity types, units of analysis, and types of innovation. Nine of 12 studies found a statistically significant and positive relation for educational diversity, four of six for occupational diversity, six of 11 for functional background diversity, four of seven for industry experience, three of five for tenure diversity, 10 of 15 gender diversity, and three of five for age diversity (with two studies finding a negative relation). As to diversity in nationality, ethnicity, country of origin, and culture, nine of 12 studies reported a positive relation with innovation (see Table 2). These positive relations are found across different units of analysis and types of innovation performance. There is a large heterogeneity in the types of innovation performance indicators applied in the analysis and a mix of units of analysis and types of diversity, which makes it difficult to compare results. Even studies that use CIS-based indicators and employer–employee linked data do not necessarily employ the same unit of analysis or type of diversity or similar types of innovation performance indicators. Thus, there is a lack of replication studies in research on workplace diversity and innovation.

The studies in our review demonstrate a relatively large number of findings that are not statistically significant. It is notable that there is a general lack of discussion of these non-significant effects and the reasons for and implications of those effects in both research and practice. We were also slightly surprised to observe relatively few articles that demonstrate a negative relation between workplace diversity and innovation performance. Only five studies found a statistically significant negative relation (Camelo et al., 2010; Østergaard et al., 2011; Galia and Zenou, 2012; Brunetta et al.,2020; Solheim et al., 2020). Although we cannot confirm this based on our findings, we must ask whether there might be a publication bias in research on workplace diversity and innovation, with studies finding negative relations less likely to have been submitted for publication.

Several studies also investigate more complex or multiple relations between workplace diversity and innovation; these articles appear several times in this review. In these instances, we observe a variety of effects between diversity constructs and innovation performance (e.g., Østergaard et al., 2011, Lee, 2014; Zouaghi et al., 2020) and that some measures of workplace diversity only have positive effects for specific types of innovation (e.g., Galia and Zenou, 2012; Özgen et al., 2013; Nathan and Lee 2013). A few cases also demonstrate curvilinear effects (Kim and Kim, 2015; González-Moreno et al., 2018; Wikhamn and Wikhamn, 2020), while others contained hypotheses predicting this effect that did not find empirical support (e.g., Østergaard et al., 2011). Studies that apply measures of related and unrelated variety, which in some ways capture whether there is too much diversity, also present mixed results (Herstad and Solheim, 2018, Herstad et al., 2019; Solheim et al., 2020). From this perspective, then, the double-edged sword argument holds. In other instances, the results demonstrate a complex relation between workplace diversity and innovation. In some cases, the effect of diversity only becomes positive when moderated by human resources practices (Chi et al., 2009), strategic consensus (Camelo et al., 2010), or communication practices (Li et al., 2016). From a theoretical perspective, these studies take diversity as a resource as a point of departure but identify the management of attitudes and group processes as important for innovation success. This emphasizes that diversity only refers to representation of a variety of knowledge resources, while practices have to ensure that the potential offered by diversity is realized in practice. In several cases, a lack of diversity can be mitigated: for example, by increased team dynamics (Wei and Lau, 2012) and external collaborations (Mohammadi et al., 2017). It also bears considering that there is a large heterogeneity of contexts, diversity constructs, and innovation performance measures in these articles.

## Measurement and Analytical Challenges

The studies on workplace diversity and innovation reveal several empirical issues. As to types of diversity measures applied, most studies rely on standardized methods; consequently, many refer to Harrison and Klein (2007), whose work elaborates on which type of diversity measure is most suitable for a given type of diversity construct. Consequently, most apply standard measures like the Blau measure (e.g., Bantel and Jackson, 1989) or Shannon's Entropy (e.g., Østergaard et al., 2011) for categorical diversity constructs, whether function, education, industry, or gender. A simple measure of shares is also applied in the case of gender (e.g., Nathan and Lee, 2013; Lee, 2014; Rejeb et al., 2020) but only when emphasizing minority groups. For diversity constructs like age and tenure, measures like standard deviation (e.g., Chi et al., 2009), or coefficient of variation (e.g., Camelo et al., 2010) tend to be used, unless age and tenure are measured in cohorts rather than specific years. Some studies deviate from these measures by using indicators like uniqueness (e.g., Özgen et al., 2013, 2017) or applying traditional Cobb–Douglas production functions (Laursen et al., 2020). More recently – and following the rise of the concept of related and unrelated variety (Frenken et al., 2007) – these measures have also been used to measure workplace diversity. As a measurement concept, related and unrelated variety attempts to incorporate some dimension of proximity and disparity and thus address limits on the extent to which different knowledge can be integrated (Cohen and Levinthal, 1990; Nooteboom, 2000). In our review, these articles exclusively focus on industry experience (Herstad and Solheim, 2018; Herstad et al., 2019; Solheim et al., 2020).

Studies that focus on diversity in culture, nationality, and ethnicity tend to use the fractionalization index, which is defined as one minus a squared sum of each cultural group. This approach is quite similar to the classical Herfindahl index to measure concentration in a market. However, this index is sensitive to the size of the dominant group of native employees. Therefore, studies often disregard the native-employee group and simply calculate the index based on the shares of non-native groups (e.g., Özgen et al., 2013). This is done to capture the variation in different nationality groups rather than the balance between natives and non-natives. The tendency to calculate diversity indices using a subset of a workforce was also seen in studies of educational diversity (e.g., Mohammedi et al., 2017). Other studies make changes to the above-mentioned diversity measures, such as normalizing the index (Bocquet et al., 2019). The operationalization of workplace diversity is likely to influence the empirical investigation of the relation between diversity and performance (Solanas et al., 2012). The choice of operationalization of diversity in terms of attributes and the choice of measure of diversity often depend on the type of data available and the definition of the type(s) and dimension(s) of diversity. However, each diversity measure varies subtly in the weight given to a number of different groups and the balance between these groups. Thus, differences in the choice of diversity indices and differences in the delimitation of groups by attributes make it difficult to compare results between seemingly similar studies. Therefore, caution should be taken when interpreting and comparing the results of different studies of workplace diversity.

As noted above, we also see a large variation in type of innovation measures (Tables 1 and 2 present an overview on these different measures). The choice of innovation performance indicator affects not only the choice of econometric models but also the interpretation of results and comparability with other studies. Table 2 also shows that studies differ by unit of analysis, ranging from small teams to large workplaces. Thus, as highlighted in the Findings section, studies vary by type of diversity, type of innovation, and unit of analysis, all of which raise concerns about the replicability of results. Furthermore, these studies also differ in their measurement of diversity, so even if the majority of studies report a positive relation between workplace diversity and innovation, these findings should be interpreted with some caution.

There are also unanswered questions related to endogeneity. Studies of workplace diversity and innovation argue that workplace diversity affects innovation performance but – as noted in several studies – the causality might move in the opposite direction. That is, workplace diversity might be a result of a firm's innovativeness, by which workers with a particular attribute are attracted to a given firm precisely because of its innovativeness. Alternatively, they might have been hired in the first place to perform innovation as a result of a firm's prior decision to focus on innovation. Some studies have tried to address endogeneity issues related to cultural and nationality diversity since the geographical distribution of non-native workers is not uniform, and some firms are therefore more likely to hire non-native workers because they are overrepresented in a local labor market (see Özgen et al., 2013; Brixy et al., 2020), but there are still many unsolved endogeneity issues.

## Future Directions

Despite extensive and heterogeneous efforts to investigate the relation between workplace diversity and innovation performance, several avenues for future research on the topic remain; they involve both conceptual and empirical extensions to existing research.

First, we suggest investigating in greater detail how diversity in different parts of an organization matters for the firm's overall innovation performance. Existing research tends to focus on diversity in either specific units or the firm's entire workforce, with only Schubert and Tavassoli (2020) and Brixy et al. (2020) explicitly addressing the difference in the innovation performance of different units. If we are to take claims that innovation is the result of an interactive process that involves all employees seriously (Lundvall, 1992; Østergaard et al., 2011), are appreciative of the mechanisms that allow innovations to emerge and are aware of how different parts of an organization can contribute differently to this process, such research efforts will provide us with more precise evidence on the nature of the mechanism(s) underlying workplace diversity and innovation performance. Future studies might investigate (i) where in the organization diversity really makes a difference and (ii) the extent to which there exist complementarities, dependencies, or substitute effects of diversity in different parts of an organization.

A natural extension of the first proposed research trajectory would be to extend efforts in how different work practices moderate and mediate the relation between workplace diversity and innovation performance. Theoretically, the variety in knowledge underlying workplace diversity is assumed to contribute to innovation performance, but we recognize that not all expected positive effects materialize in practice. In our discussion of the two overarching theoretical perspectives, we have already offered an argument as to why this might occur: workplace diversity not only links to a particular richness in knowledge and resources but also affects worker attitudes and group processes. Some articles in our review demonstrate this complex interaction, as we see moderators that play a key role in how individuals interact and how diversity in general is accepted (Chi et al., 2009; Camelo, 2010). This opens a line of inquiry that focuses more on organizational features and human resources practices in firms. Indeed, openness to diversity has been shown to have a positive effect on innovation performance (Østergaard et al., 2011), but clear measures are lacking in the literature. Another issue that might be addressed is how work organization could be an important moderator that enables the benefits of diversity to materialize. Similarly, paying specific attention to recruitment procedures and onboarding process of new (diverse) employees might drive the successful integration of diversity into an organization. This also relates to the debate that diversity is not necessarily equal to inclusion in the workplace. This trajectory and being more explicit on the role of diversity in different parts of the organization could provide managers with insights into the measures to take to ensure that diversity's benefits are realized and possible drawbacks mitigated.

Third, and in close alignment with the theme of this handbook on spatial diversity and business economics, a systems perspective might be adopted to investigate the interaction between workplace organization and diversity in the environment in which the organization is embedded. Few studies have sought to disentangle regional from firm-level effects. When it comes specifically to diversity and innovation, we could only identify Lee (2014), who investigates how cultural diversity at the regional level and firm-owner cultural diversity relate to innovation performance. Such embeddedness could influence a firm's ability and need to become diverse to strengthen innovation performance. First, local labor markets necessarily affect a firm's ability to attract a diverse work force. Large and rich labor markets offer a plentiful labor pool from which firms can recruit and build a more diverse workforce. On the other hand, peripheral areas with thin labor markets simply do not offer the same opportunities. Alternatively, one might question the extent to which internal workforce diversity is necessary, depending on the diversity in the broader labor market. Firms located in rich and diverse regions might be able to draw on diverse knowledge from their environment, which makes internal diversity a less pressing issue; in other words, there would be a substitution effect. Mohammadi et al. (2017) argue for a substitution effect between diversity and collaboration; in this setting, it would be extended to labor markets in general.

The fourth avenue of future research points more explicitly to addressing the measurement and analytical challenges that we raised in the previous section. We observed a large variety in innovation performance measures and diversity constructs, which hampers comparability between studies and demands caution in the interpretation of results. Some of these variations in diversity constructs are likely to influence empirical results, but this issue was rarely addressed in the papers. Thus, future research could address the econometric implications of choice of diversity construct and measurement. Furthermore, future research on workplace diversity and innovation should report the sensitivity of the choice of diversity construct and measure by, for example, using different diversity measures. In addition, as raised in the Findings section, future studies need to include a discussion of effects that were found but were not statistically significant. Many studies in our review have several such findings, which may be explained by the empirical approach, the data, a combination of the two, or something else entirely. We found a tendency to focus on statistically significant results and largely ignore the frequent non-significant results. Causality is difficult to establish since most of the articles rely on a cross-sectional research design. While register data offers opportunities to apply a panel data methodology, this approach has not been used to a meaningful extent. Since patent data and CIS (and CIS-style) data are gathered at regular intervals, it is possible to follow the composition of the workforce and innovation performance over time. Thus, several questions related to endogeneity concerns remain unanswered. Finally, we observe a large heterogeneity in the studies in context, diversity construct, and innovation performance. To establish more firmly how workplace diversity is related to innovation performance, replication studies are required. Replications offer opportunities to conduct systemic reviews and meta-studies that can offer powerful evidence on the relation between workplace diversity and innovation performance.

Table 1: Overview of papers on workplace diversity and innovation performance

| | Author | Unit of analysis | Type of innovation | Types of diversity | Nature of the relation | Exp. dir. of the relation | Data type | Context | Main results |
|---|---|---|---|---|---|---|---|---|---|
| 1 | Bantel and Jackson (1989) | TMT | Technological and administrative innovations. | Age, tenure, education, and functional background. | Direct | Positive and negative for age and tenure, positive for education and functional | Survey | 199 US Banks | Non-significant effect for tenure, age, and education, diversity on innovation performance, but positive for functional diversity |
| 2 | Van der Vegt and Janssen (2003) | Work Teams | Innovative work behaviour (searching, promotion and realization) | Age, gender, ethnicity, and cognitive | Moderated by perceived task and goal interdependence. | Positive | Survey | 41 work teams in a Dutch financial firm | Positive relation between individual innovative behaviour and perceived task interdependence for high demographic and cognitive group diversity. |
| 3 | Yap et al. (2005) | TMT | Number of product and process innovation. | Functional, age, and tenure. | Direct (moderated by firm size) | Positive | Survey | 50 SME in Singapore | Positive relation for intrapersonal functional diversity, but negatively moderated for large firm. Negative for age diversity. Non-significant for tenure diversity. |
| 4 | Chi et al. (2009) | R&D team | Processes, products, or procedures that are new to the team | Tenure, educational, and industry experience. | Direct (moderated by HR practices) | Curvilinear | Survey | 67 R&D teams from 35 Taiwanese high-tech firms | Curvilinear relationship between tenure diversity and innovation performance. Education diversity is not significant. When education diversity is moderated by HR practices education diversity is positive. |
| 5 | Camelo et al. (2010) | TMT | Number of new products and number of improved products | Tenure, and functional. | Direct (moderated by strategic consensus) | Positive | Survey | 97 firms from innovative sectors in Spain | Tenure and functional diversity are both negatively related of innovation performance. But with the presence of strategic consensus, the educational diversity measure turns positive. |
| 6 | Østergaard et al. (2011) | Workforce | Product and service innovation | Gender, country of origin, age, education, and diversity policy. | Direct | Curvilinear, negative/neutral for age diversity | Survey and register data | 1648 Danish firms | Positive for educational diversity and gender diversity, but negative for age diversity. Ethnic diversity is not significant. No curvilinear effect detected. Positive relation between diversity policy and innovation performance. |
| 7 | Galia and Zenou (2012) | Board | CIS-based product, process, organisational, marketing innovation | Gender, age. | Direct | Positive. Negative for age diversity related to organizational innovation | Survey and annual reports | 176 French firms | Gender diversity leads to more marketing innovation while not significant for organizational innovation. Age diversity is positive for product innovation but negative for organisational innovation. Age diversity is not significant for marketing innovation. |
| 8 | Wei and Lau (2012) | TMT | Composite measure (NPD, R&D budget and technological innovation) | Age, education, function, and tenure. | Direct and moderated by team dynamics | Positive/negative | Survey | 600 Chinese firms | Positive for age, tenure, and educational diversity. Team dynamics increases innovation performance when diversity is low. |
| 9 | Díaz-Garcia et al. (2013) | R&D workers | Product and process innovation | Gender | Direct | Positive | PITEC survey | 4277 Spanish firms | Gender diversity is positive and significant related to innovation performance. |
| 10 | Somech and Drach-Zahavy (2013) | Teams | Team creativity: generation of ideas that are both novel and useful to the team | Functional, gender, education, and function. | Indirect (functional promote creativity that support innovation) | Positive | Survey and interview | 96 Israeli primary care teams in a health maintenance organisation | Diversity measures have a positive and significant relation with team creativity. While team creativity has a positive effect on innovation performance, but only in an innovative climate. |
| 11 | Özgen et al. (2013) | Workforce | CIS-Based innovation activity, product and process | Cultural (nationality) | Direct | Positive | CIS survey and register data | 4582 Dutch firms | Cultural diversity has a positive effect on innovation activity and product innovation. Cultural diversity has no significant relation with process innovation |
| 12 | Nathan and Lee (2013) | TMT (owners, partners) | New products and services, process innovation commercialization of innovation | Ethnic, migrant | Direct | Positive and negative | Survey | 7600 London-based private firms | There is a small but robust positive relation for migrant and ethnic diversity bonus on innovation performance for product and process innovation. When focusing on commercialization, most of the diversity measures are not significant. |
| 13 | Parrotta et al. (2014) | Workforce | Patents | Cultural/nationality, educational, and demographic. | Direct (but differences for white and blue collared workers). | Positive | Register and patent data | 12000 Danish firms | Positive and significant for cultural diversity for both white and blue collared workers, but stronger for white collared workers. No significant effect on educational and demographic diversity. |
| 14 | Lee (2015) | TMT (owners, partners) | Product and process innovation | Ethnic, migrant | Direct | Positive (but highlight the mixed results in research) | Survey | 200 UK SMEs | Positive relation between share of migrant owners and partners and product and process innovation. There are however, diminishing effects. Effect of ethnic diversity is not significant. |
| 15 | Kim and Kim (2015) | Board | Application in a new-to-the-firm patent class | Functional, occupational, and relational. Combined in one measure of board capital diversity. | Direct (moderated by CEO ownership and board ownership) | Curvilinear | Korea Investors Services database and patent data | 108 listed Korean manufacturing firms in R&D intensive industries | Finds an inverted u-shaped relationship between board capital and innovation performance. This relation is positively moderate by CEO ownership, but not with board ownership. |
| 16 | Fernández (2015) | R&D workers | Products, services and process innovation | Gender and functional. | Direct | Positive (insignificant for foreign firms) | PITEC survey | 30327 Spanish firms | Gender and functional diversity is positive related to innovation performance, especially product innovation. There is an inverted u-shape for gender diversity. Gender diversity is, however, non-significant for foreign firms. |
| 17 | Kristinsson et al. (2016) | Founding team | Generation of ideas and implementation of these | Informational. | Direct (moderated by causation logic) | Positive | Survey | 133 new technology-based ventures | Positive and significant relation between informational diversity and idea generation and implementation. Positive moderation by causation logic for idea generation negative for implementation. |
| 18 | Cheung et al. (2016) | R&D team | Introducing new service, methods and procedure | Functional | Indirect mediated by knowledge sharing (moderated by affect-based trust). | Positive | Survey | 117 teams in a Chinese IT firm | Finds a non-significant indirect negative relation between functional diversity and innovation performance. Affect based trust made the negative relation less negative. |
| 19 | Li et al. (2016) | TMT | Ambidextrous innovation | Functional and education | Direct and indirect mediated by TMT level debate and decision - making comprehensiveness | Positive | Survey and interview | 179 listed Chinese high-tech firms | TMT diversity only has an indirect effect on innovation performance through TMT debate and comprehensive decision-making. No direct relation with innovation performance |
| 20 | Ruiz- Jiménez et al. (2016) | TMT | Four indicators of innovation | Gender | Indirect (gender diversity mediates knowledge combination capability) | Positive | Survey and interview | 205 Spanish tech-based SME | Direct effects are not significant but when interacted with knowledge combination gender diversity is positively associated with innovation performance. |
| 21 | Protogerou et al. (2017) | Founding Team | Introduction of new product, and radicalness of innovation | Functional and occupation. | Direct | Positive | Survey | 3962 young KIBS in 10 Europeans countries | Positive and significant relation between functional diversity innovation performance. Occupational diversity turns out to be not significant. |
| 22 | Tzabbar and Margolis (2017) | Founding team | break through innovations (patent citations) | Education | Direct and moderated by growth stage | Positive | Patent data and various database s | 578 biotech firms in US | Positive and significant relation between educational diversity and break through innovation. This effect is stronger in the growth stage of the business. |
| 23 | Garcia Martinez et al. (2017) | R&D workers | Product and process innovation | Gender, education, and skills. | Direct | Positive | PITEC survey | 1000+ Spanish firms | Positive and significant relation between gender, education and skill diversity and innovation performance. Interaction between the different diversity measures demonstrate negative signs. |
| 24 | Mohammadi et al. (2017) | Workforce | Sales from radical innovation | Ethnic, education | Direct | Positive | CIS survey and register data | 3888 Swedish firms | Ethnic and education diversity is positively associated with innovation performance. Negative interaction between search breadth and workforce diversity. |
| 25 | Özgen et al. (2017) | Workforce | Product and process innovation | Culture/nationalit y | Direct and indirect | Positive/negative | CIS survey and register data | 4931 Dutch firms | Cultural diversity is positively related to process innovations in some specifications. Generally insignificant results. |
| 26 | González-Moreno et al. (2018) | R&D workers | Product innovation | Gender | Direct | Inverted U-shape | PITEC survey | 3540 Spanish manufacturing | Finds an inverted u-shape relationship between gender diversity and product innovation. |
| 27 | Solheim and Herstad (2018) | Workforce | Innovation and patents | Industry experience, and education (related and unrelated). | Direct | Positive | CIS survey and register data | 2942 Norwegian firms | Related experience variety is positive and significant for innovation in some model specifications. Unrelated experience variety is positive and significant for patenting in some model specifications |
| 28 | Rejeb et al. (2020) | Board | Ambidextrous innovation | Gender | Moderator | Positive for relation between strategy and innovation, negative for relation between control and innovation | Survey | 81 listed Tunesian firms | Non-significant effect for relation between control and innovation. Significant and positive for relation between strategy and innovation |
| 29 | Li (2019) | Board | Patent: Entering into a new technological space | Industry experience | Direct | Positive | Patent data, compust at, boardex, | 895 US listed firms with patenting activities | Significant and positive relation between industry experience diversity and innovation performance |
| 30 | Moser et al. (2019) | Teams | Team innovations were rated using a five-point scale on four dimensions: magnitude, radicalness, novelty, and impact" | Occupation | Moderator (mediate information sharing and helping behaviour) | Positive | Survey and interview | 185 UK health care teams | Direct effect of occupational diversity on innovation performance is not significant. There is, however, a positive and significant interactions between occupational diversity and information sharing and helping behavior on innovation performance. |
| 31 | Boone et al. (2019) | TMT | Patent Count | Nationality, experience, functional | Indirect (nationality diversity is positive related to corporate entrepreneurship, which is positively related to innovation) | Positive | OECD data plus patent statistics and annual reports | Panel of 165 multinational firms from 20 OECD countries | Positive and significant relation between nationality diversity and the number of patents. Experience and functional diversity are not significant. |
| 32 | Li and Huang (2019) | TMT | Patent count | Education and tenure. | Moderator | Positive | Data from patent statistics and annual reports. | 283 patenting firms in Taiwan | Moderating effect of tenure diversity on R&D investment and moderating effect of educational diversity on international diversification is supported. Other moderating effects are not supported. Positive significant direct effect of tenure diversity. |
| 33 | Bocquet et al. (2019) | Workforce | Technological (process or product) innovation" | Gender and nationality. | Direct (strategic CSR positively relates to diversity) | Positive | Survey | 1348 SMEs from Luxembourg | Non-significant positive effect of gender diversity. Positive and significant (5%) of diversity in nationality on innovation performance. |
| 34 | Herstad et al. (2019) | Workforce | Product innovation | Industry experience (related and unrelated). | Direct | Positive | CIS Survey and register data | 1424 Norwegian firms | Unrelated variety in industry experience is positively and significant related to being innovation active. Related variety in industry experience is positive and significant related for product innovation, while unrelated variety in industry experience is not significant. |
| 35 | Schubert and Tavassoli (2020) | TMT and Middle Management Team | Innovation engagement and product innovation | Education, gender, and nationality. | Direct | Positive for education | CIS survey and register data | 486 Swedish firms | Positive relation between education diversity in TMT and innovation engagement. Positive and significant relation between MMT education diversity and product innovation and new to the market innovation. All other specifications and diversity measures are not significant |
| 36 | Brixy et al. (2020) | founding team, and workforce | Product and process innovation | National origin and origin compared to the region. | Direct | Positive | Survey and register | 3293 German startups | National diversity among founders and employees is significant and positive on innovation performance when applying unusualness measure. In other cases, the measures are not significant. |
| 37 | Brunetta et al. (2020) | R&D teams | Completed clinical trials | Institutional identities. | Direct (moderated by duration) | Negative | Data from clinical trials register | 3658 clinical trials | Institutional diversity is significant and negative related to innovation performance. But the longer they have collaborated, the less negative the relation becomes. |
| 38 | Xie et al. (2020) | R&D teams | Share of sales based on new products / R&D efficiency | Gender | Direct (moderated by organisational level factors) | Positive | Survey | 18217 Chinese firms | Significant and positive between gender diversity and innovation performance. Some positive and significant interactions with organisational level factors. |
| 39 | Zouaghi et al. (2020) | R&D teams | CIS-based. Product and process innovation | Gender, education, and skill | Direct and moderator | positive, but negative interactions with other diversity measures | PITEC survey | 30999 Spanish firms | Diversity variables (gender, skills, and education) are positively associated with product and process innovations. Negative interaction between diversity in gender and skills diversity as well as gender and education diversity. |
| 40 | Solheim et al. (2020) | Workforce | CIS-based. product innovation | Industry experience (related unrelated). | Direct | Positive | CIS survey and register data | 1463 Norwegian manufacturing firms | No support for collective diversity experience on innovation performance (neither radical nor incremental), but positive for related and unrelated variety 39on incremental and radical innovation in some model specifications |
| 41 | Laursen et al. (2020) | workforce (high-skilled) | Citation-weighted patent count | Cultural. | Direct | Positive | Register combine d with survey and patent data | 16241 Patenting and R&D active Dutch firms | Positive and significant relation between cultural diversity among high skill workers and innovation performance. |
| 42 | Wikhamn and Wikhamn (2020) | R&D workers(R&D) | Share of sales based on new products | Gender. | Direct | Positive/Negative | CIS and R&D survey | 1114 Swedish firm | positive non-linear, upright u-shape relationship between gender diversity and innovation performance. |

Table 2: Overview of findings on the relation between different types of diversity and innovation[3]

| | | Product and service innovation | | | Process innovation | | | Other types of innovation | | | Patent based indicators | | | Radical innovation | | |
|---|---|---|---|---|---|---|---|---|---|---|---|---|---|---|---|---|
| **Type of Diversity** | **Unit of analysis** | **Main result on the relation between innovation and diversity: positive (+)/negative(-)/ not statistically significant (0)** | | | | | | | | | | | | | | |
| | | **+** | **-** | **0** | **+** | **-** | **0** | **+** | **-** | **0** | **+** | **-** | **0** | **+** | **-** | **0** |
| **Education diversity (papers: 1, 4, 6, 8, 10, 13, 23, 24, 27, 32, 35, 39)** | | | | | | | | | | | | | | | | |
| | Workforce | 6 | | | | | | | | 27 | 13;27 | | 13;27 | 24 | | |
| | TMT/board/founders | | | 35 | | | | 8 | | 1 | 22 | | 32 | | | 35 |
| | R&D workers/teams | 39 | | | 39 | | | 23 | | 4;23 | | | | 23 | | |
| | Teams | 35 | | | | | | | | 10 | | | | 35 | | |
| **Occupation/skills diversity (papers: 15, 19, 21, 23, 30, 39)** | | | | | | | | | | | | | | | | |
| | Workforce | | | | | | | | | | | | | | | |
| | TMT/board/founders | | | | | | | | | 19 | 15 | | | | | |
| | R&D workers/teams | 39 | | 21 | 39 | | | 23 | | | | | | 23 | | 21;23 |
| | Teams | | | | | | | 30 | | | | | | | | |
| **Functional background diversity (papers: 1, 3, 5, 8, 10, 15, 16, 18, 19, 21, 31)** | | | | | | | | | | | | | | | | |
| | Workforce | | | | | | | | | | | | | | | |
| | TMT/board/founders | 21 | 5 | | | | | 1;3 | | 8;19 | 15 | | 31 | 21 | | |
| | R&D | 16 | | | 16 | | | | | 18 | | | | | | |
| | Teams | | | | | | | 10 | | | | | | | | |
| **Industry experience diversity (papers: 4, 27, 29, 31, 34, 37, 40)** | | | | | | | | | | | | | | | | |
| | Workforce | 34;40 | 40 | 34;40 | | | | 27 | | 27 | 27 | | 27 | 40 | | 40 |
| | TMT/board/founders† | | | | | | | | | | 29 | | 31 | | | |
| | R&D‡ workers/teams | | | | | | | | 37 | 4 | | | | | | |
| | Teams | | | | | | | | | | | | | | | |
| **Tenure diversity (papers: 3, 4, 5, 8, 32)** | | | | | | | | | | | | | | | | |
| | Workforce | | | | | | | | | | | | | | | |
| | TMT/board/founders | | 5 | | | | | 8 | 5 | 3 | 32 | | | | | |
| | R&D workers/teams | | | | | | | 4 | | | | | | | | |
| | Teams | | | | | | | | | | | | | | | |
| **Gender diversity (papers: 6, 7, 9, 10, 13, 16, 20, 23, 26, 28, 33, 35, 38, 39, 42)** | | | | | | | | | | | | | | | | |
| | Workforce | 6 | | | | | | | | 33 | 13 | | 13 | | | |
| | TMT/board/founders | | 7 | 20;35 | | | 7 | 7;28 | | 7 | | | | | | 35 |
| | R&D workers/teams | 16;26;38;39;42 | | | 16;39 | | | 9;23 | | 23 | | | | 9;23 | | |
| | Teams | | | 35 | | | | | | 10 | | | | | | 35 |
| **Age diversity (papers: 1, 6, 7, 8, 13)** | | | | | | | | | | | | | | | | |
| | Workforce | | 6 | | | | | | | | 13 | | 13 | | | |
| | TMT/board/founders | 7 | | | | | 7 | 8 | 7 | 1;7 | | | | | | |
| | R&D workers/teams | | | | | | | | | | | | | | | |
| | Teams | | | | | | | | | | | | | | | |
| **Nationality, Ethnicity, Country of Origin, Cultural diversity (papers:6, 11, 12,13,14, 24, 25, 31, 33, 35, 36, 41)** | | | | | | | | | | | | | | | | |
| | Workforce | 11 | | 6;25 | 25 | | 11;25 | 11;33;36 | | | 13;41 | | | 24 | | |
| | TMT/board/founders | 12;14 | 14 | 35 | 12;14 | 14 | | 36 | 12 | | 31 | | 31 | | | 35 |
| | R&D workers/teams | | | | | | | | | | | | | | | |
| | Teams | | | 35 | | | | | | | | | | | | 35 |
| **Informational diversity (paper:17)** | | | | | | | | | | | | | | | | |
| | TMT/board/founders | | | | | | | 17 | | | | | | | | |
| **Demographic composite diversity measure – age, gender, and ethnicity (paper: 2)** | | | | | | | | | | | | | | | | |
| | Teams | | | | | | | | | 2 | | | | | | |
| **Cognitive group diversity (paper: 2)** | | | | | | | | | | | | | | | | |
| | Teams | | | | | | | | | 2 | | | | | | |

†: including work experience and industry background

‡: Including institutional diversity

[3] Numbers refer to the paper numbers in Table 1